\documentclass[10pt,conference]{IEEEtran}
\IEEEoverridecommandlockouts
\usepackage[numbers]{natbib}
\usepackage{pifont}
\usepackage{amsmath,amssymb,amsfonts}
\usepackage[table]{xcolor}
\usepackage{graphicx}
\usepackage{url}
\usepackage{multirow}
\usepackage{verbatim}
\usepackage{dblfloatfix}    

\AtBeginDocument{%
  \providecommand\BibTeX{{%
    \normalfont B\kern-0.5em{\scshape i\kern-0.25em b}\kern-0.8em\TeX}}}

\newif\ifdraft
\draftfalse
\def\boldification #1 {\ifdraft\textbf{#1\newline\indent}\else\relax\fi}

\begin{document}

\title{Please Don't Go - Increasing Women's Participation in Open Source Software}

\author{\IEEEauthorblockN{Bianca Trinkenreich}
\IEEEauthorblockA{\textit{Northern of Arizona University} \\
\textit{Flagstaff, AZ, USA} \\
bt473@nau.edu}
}

\maketitle

\thispagestyle{plain}
\pagestyle{plain}

\begin{abstract}

Women represent less than 24\% of the software development industry and suffer from various types of prejudice and biases. In Open Source Software projects, despite a variety of efforts to increase diversity and multi-gendered participation, women are even more underrepresented (less than 10\%). My research focuses on answering the question: How can OSS communities increase women's participation in OSS projects? I will identify the different OSS career pathways, and develop a holistic view of women's motivations to join or leave OSS, along with their definitions of success. Based on this empirical investigation, I will work together with the Linux Foundation to design attraction and retention strategies focused on women. Before and after implementing the strategies, I will conduct empirical studies to evaluate the state of the practice and understand the implications of the strategies.

\end{abstract}


\section{Research Problem and Motivation}

Gender diversity has positive effects on productivity, bringing together different perspectives; improving outcomes \cite{vasilescu2015gender}, innovation, and problem-solving capacity; and leading to a healthier work environment \cite{earley2000creating}. A diverse development team is more likely to properly comprehend users’ needs and contribute to a better alignment between the delivered software and its intended customers \cite{muller1993participatory}. Despite this, women represent only 5.2\% of contributors in Apache Software Foundation \cite{asf2016survey}, one of the largest and best-known OSS communities. More broadly, women represent 10.35\% of OSS contributors \cite{robles2016women}, only 9\% of GitHub users \cite{vasilescu2015data}, and author less than 10\% of contributions \cite{bosu2019diversity,zacchiroli2020gender}. I will investigate women's participation in OSS, including their motivations to join, what attracts them to OSS, the pathways they follow as they evolve on a project (or to achieve their perceptions of success), the challenges they face and their motivations to stay, take breaks or leave, what retain or repeal them from OSS. Ultimately, the overall goal of this project is to help OSS projects devise strategies to attract and retain women, while helping these women contributors to attain their own goals. 

\section{Background and Related Work}
\label{sec2}

Strategies suggested by previous work to attract and retain women include issuing code of conduct statement \cite{tourani2017code,lee2019floss,fossatti2020gender}; adopting feminist and social justice principles \cite{fossatti2020gender}, promoting women to leadership positions \cite{bosu2019diversity},
providing places for women to build their leadership capacity and to engage in the development of the community with
norms and values consistent with their own vision \cite{qiu2010joining}, 
focusing on the first social experiences through programs such as mentorships \cite{kuechler2012gender}, and fixing the gender-bias issues in non-inclusive tools and infrastructure \cite{mendez2018open}. Some strategies that are discouraged by the literature include setting quotas for women. Just increasing the proportion of women can lead to questioning stereotypes and decisions to simply re-classify types of work that are currently packaged in masculine-feminine stereotyped specialties. Another discouraged strategy is related to coding schools that train women in specialties where they are already represented. These schools might perpetuate the disadvantage of women by their femaleness of behavior, and contribute to gender homophily by creating more women-to-women ties among the participants. \cite{vedres2019gendered}.

\section{Approach and Uniqueness}

\begin{figure*}[hbt]
     \centering
     \includegraphics[width=1\textwidth]{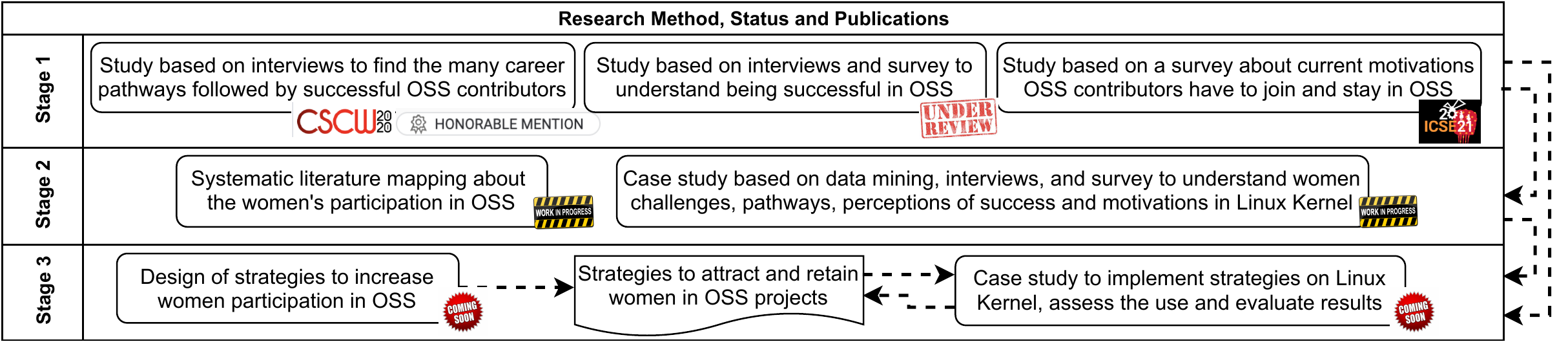}
     \caption{Research Design}
     \label{fig:research_design}
 \end{figure*}

\begin{figure}[hbt]
     \centering
     \includegraphics[width=0.38\textwidth]{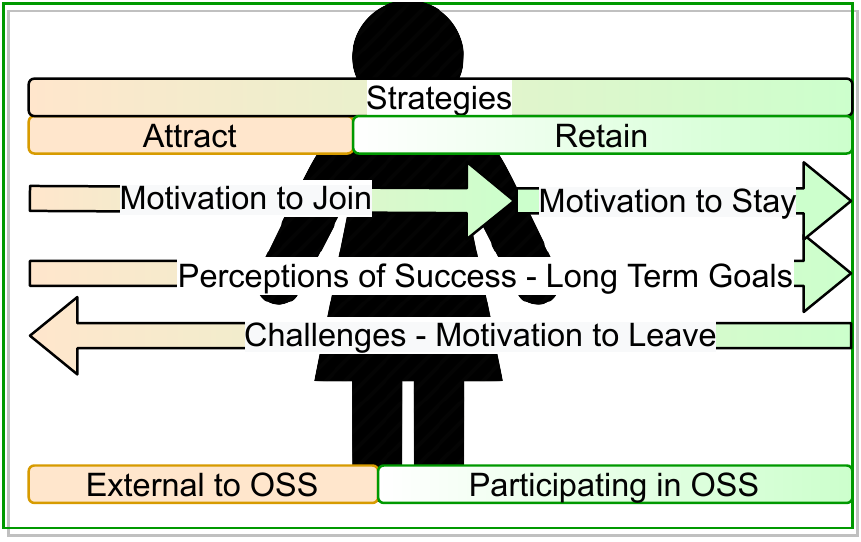}
     \caption{Strategies focused on the forces to join, stay, and leave OSS}
     \label{fig:forces}
 \end{figure}

My research comprises three stages and adopts mixed methods to accomplish its goal, as depicted in Fig. \ref{fig:research_design}.
I have already finished Stage 1 and have begun Stage 2, exploring the career pathways, goals, definitions of success, and motivations that influence the decisions of a contributor to join, stay, take breaks, or challenges that can drive a contributor to leave an OSS project. The studies include all genders as I want to compare findings between genders. My hypothesis is that an OSS community or organization can better attract and retain contributors when considering both their motivations and the multitude of factors that underpin their definition of success. For this purpose, I will create strategies to 1. attract and retain women based on the forces compelling them to join and stay in OSS, and 2. diminish the forces compelling them to drop out of OSS, as depicted in Fig. \ref{fig:forces}.

In our previous studies about contributors' motivations for joining and staying in OSS~\cite{gerosa2021motivation} and their perceptions of what it means to be successful in OSS (currently under peer review), I observed that while motivation and success perceptions are interrelated and complement each other, they are forces that can play different roles. An example is one of our participants, who reported being motivated to join by “reputation”, but who perceived success as “getting paid to contribute”. Individuals with diverse backgrounds and understandings of success may need different engagement strategies \cite{trinkenreich2020pathways}. By knowing that success is polyvalent in OSS, communities can leverage different definitions of success to support the growth of diverse individuals. For contributors who consider success as "Having contacts in several different communities", communities can promote meetups to help increase social capital. 
I aim to hear from women, as the target population who will be benefit from my study, what motivates them, how they perceive success, and the challenges they face. All empirical data will be collected from current contributors and those who left Linux Kernel, and strategies will be modeled based on holistic comprehension of motivation, success and challenges. The strategies will be applied and evaluated by Linux Kernel community.

\section{Results and Contributions}

I have been investigating the many career pathways that successful contributors from several communities follow in OSS. This study was published in a paper \cite{trinkenreich2020pathways} (CSCW 2020, honorable mention award). I have also been investigating motivations to join and stay in OSS \cite{gerosa2021motivation} (ICSE 2021, main research track) and perceptions of success in OSS. Besides extending and integrating these preliminary results, I am conducting a deeper investigation in a case study (Linux Kernel). I will mine software repositories and use a survey to understand the challenges, pathways, perceptions of success, and motivations that women face in Linux Kernel. I will also interview contributors to Linux Kernel, who will be recruited with the help of community managers supporting me on this project. I will qualitatively analyze data using coding procedures and compare different genders on their motivations to join, stay, take breaks and leave; categories of challenges they face; how people evolve and the pathways they take; how they perceive success; and their suggestions for improving inclusivity. I will also collect information about the results of the inclusivity programs that the Linux Foundation has already implemented. 

Based on my findings, I will design actionable strategies to increase women's participation in Linux Kernel in collaboration with the Linux Foundation. During the execution of the strategies, I will interact with and collect feedback from the contributors and the community managers to evaluate, learn, and improve the strategies. Additionally, I will run post-study debriefing interview sessions with women to collect their impressions, the positive and negative points of which will be used as lessons learned. I will repeat the mining and measure the observed groups after the strategies are implemented to compare the number of women in groups before and after the study and evaluate whether the strategies helped to increase women's participation.
To the best of my knowledge, there is no work that provides strategies to increase women's participation in OSS based on the factors of motivations to join, stay, and/or leave; perceptions of success; and challenges--I call these ``women's aspirations when it comes to OSS.'' This study is also novel for offering an otherwise nonexistent guideline of career pathways for women in OSS, including different roles for them to find the different pathways they can contribute and achieve their goals and versions of success.

\section{Expected Contributions}

The theoretical contribution will be multi-fold, identifying: the pathways that can be followed by OSS contributors; women's motivations to join (or not), stay, take breaks, and leave OSS projects; women's multi-faceted definitions of success; the current challenges women face; women's advice for other women; and women's suggestions to make OSS projects more inclusive. The practical contributions include guidelines for women seeking a career in OSS, highlighting the many roles, activities, backgrounds, and necessary skills. 
I will also provide actionable mechanisms for OSS projects to encourage women to join and keep contributing. The strategies will be implemented in a large OSS project that seeks to increase women's participation (Linux Kernel) and evaluated in practice. The research methodology can be used by other researchers to attract and retain other minorities besides women, and to other domains besides IT.

\bibliographystyle{IEEEtran}
\bibliography{paper}

\end{document}